\documentclass[aps,pre,english,showpacs,showkeys,preprintnumbers,twocolumn,floatfix,nofootinbib,10pt]{revtex4-2}
\usepackage{eurosym}
\usepackage{amssymb}
\usepackage{graphicx,color}
\usepackage{epsfig}
\usepackage{rotating}
\usepackage[T1]{fontenc}
\usepackage{latexsym,epsfig}
\usepackage[latin9]{inputenc}
\usepackage{amsfonts}
\usepackage{dcolumn}
\usepackage{bm}
\usepackage{babel}
\usepackage{hyperref}
\usepackage{amsmath,mathrsfs}

\begin{document}

\title{Random matrices theory elucidates the critical nonequilibrium phenomena }
\author{Roberto da Silva}
\address{Instituto de F{\'i}sica, Universidade Federal do Rio Grande do Sul,
Av. Bento Gon{\c{c}}alves, 9500 - CEP 91501-970, Porto Alegre, Rio Grande do
Sul, Brazil}

\begin{abstract}
The earlier times of evolution of a magnetic system contain more information
than we can imagine. Capturing correlation matrices $G$ of different time
evolutions of a simple testbed spin system, as the Ising model, we analyzed
the density of eigenvalues of $G^{T}G$ for different temperatures. We
observe a transition of the shape of the distribution that presents a gap of
eigenvalues from critical temperature with a continuous migration to the
Marchenko-Pastur law for the paramagnetic phase. We consider the analysis a
promising method to be applied in other spin systems to characterize phase
transitions. Our approach is different from alternatives in the literature
since it uses the magnetization matrix and not the spatial matrix of spins.
\end{abstract}

\maketitle


\textbf{1 - Introduction}: The study of the phase transitions and critical 
\cite{Stanley1971,Stanley1999} phenomena cover a considerable part of the
studies in Statistical Physics given its importance that goes beyond the
physics of the so-called \textquotedblleft natural\textquotedblright\
phenomena, approaching also the physics on the economy, society, and
biological systems \cite{Bouchaud2000,Castellano2009,Sung2018}.

Studies about the criticality of a system are intensely concentrated on the
equilibrium regime or, particularly for systems without defined Hamiltonian,
in its steady-state.

However, many authors have supported dynamic scaling laws for systems far
from equilibrium. In this case, they consider the relaxation of systems,
initially at an infinite temperature, suddenly placed at critical
temperature \cite{Janssen1989,Janssen1994,Zheng1998,Albano}, or for
nonequilibrium phase transitions of nonequilibrium models, the initial
regime out of steady-state \cite{Hinchsen,Dickman}.

Nevertheless, how important are the earlier times of a spin system's
evolution to respond to phase transitions of the systems? Based on this
question, we intend to go beyond asking how the spectral properties of
statistical mechanics systems can be affected, or more precisely, if the
traces of criticality in the earlier times can influence the spectral
properties of suitable correlation matrices defined on the systems.

Definitively, the question is fundamental since it explores if the spectral
properties can capture nuances that are not only intrinsically linked to the
steady-state of such systems.

In this case, what kind of correlation matrices are enjoyable to perform in
this study? Fortunately, we will answer these points in this paper, showing
a method that allows working with small systems compared with those
traditionally used by extrapolating systems to a thermodynamic limit.

\begin{figure}[t!]
\begin{center}
\includegraphics[width=1.0\columnwidth]{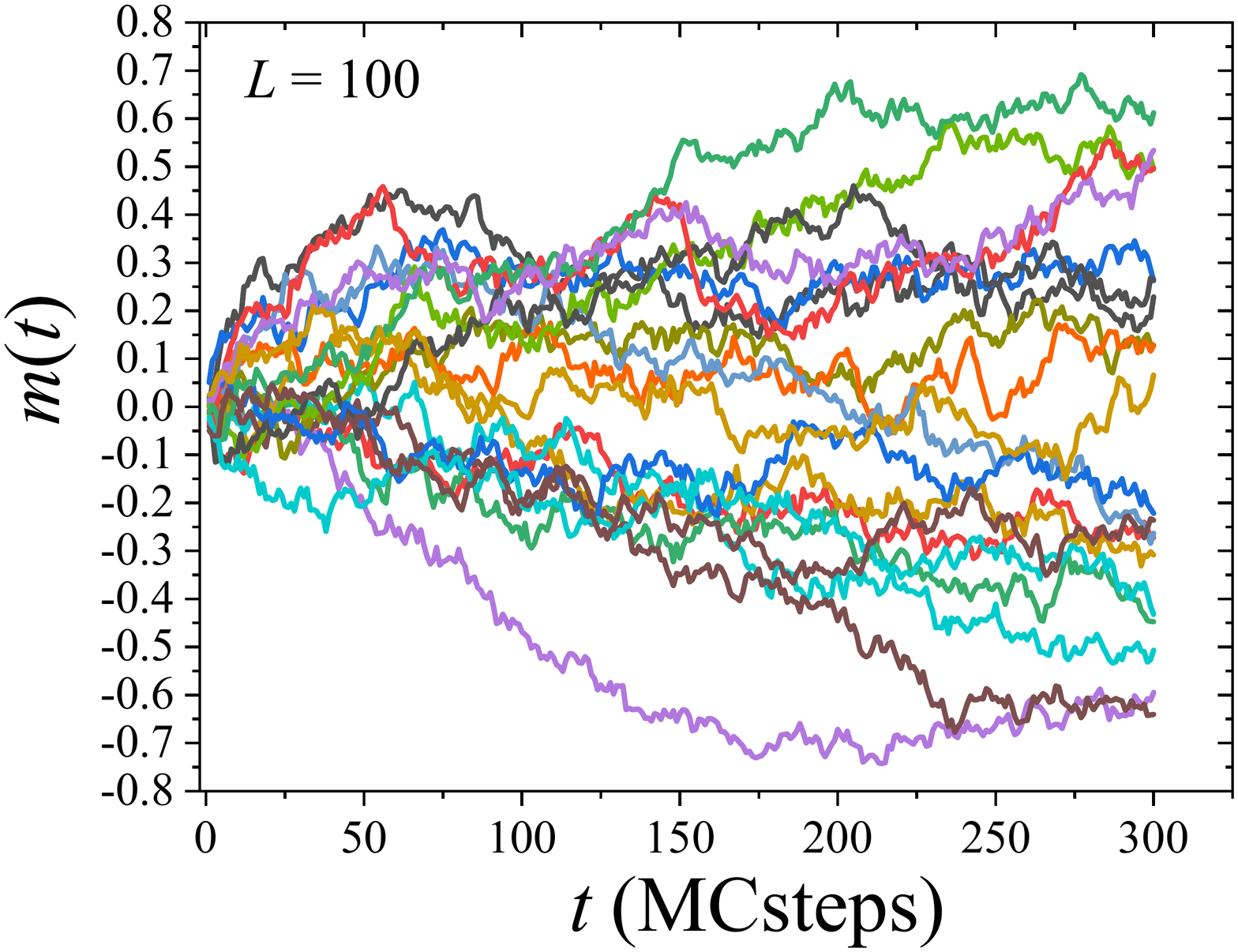}
\end{center}
\caption{An example of different time evolutions of magnetization used to
build the magnetization matrix $M$. }
\label{Fig:Differenttimeevolutions}
\end{figure}

Firstly, we know that spectral properties have a vital role in describing
and characterizing physical systems from a general point of view. For
example, in the context of random matrices, E. Wigner was the first to
observe that the distribution of eigenvalues of symmetric/hermitian
matrices, under well-behaved random entries \cite{Wigner,Mehta}, could
describe the energy spectra of the heavy atomic nucleus.

In an exciting application of random matrices, Stanley and collaborators 
\cite{Stanley,Stanley2}, using the known approach developed by Marcenko and
Pastur \cite{Marcenko,Sengupta}, showed that deviates from the bulk of
spectra of random correlation matrices built with financial market assets
are related to genuine correlations from Stock Market

Recently, some authors \cite{Vinayak2014} interestingly investigated
spectral properties of correlation matrices in near-equilibrium phase
transitions. In this case, they studied correlation matrices of the $N=L^{2}$
spins of the Ising model in the two-dimensional lattice under $\tau $ time
steps of evolution to evidence the power-law spatial correlations at a phase
transition display.

Similarly, \cite{Biswas2017} explored similar results in the steady-state
for the correlation matrix of the asymmetric simple exclusion process.
However, we believe that information about phase transitions in spin systems
is still more "primitive" than we can imagine. The traces of the phase
transition should reflect in properties of random matrices built from time
evolutions simulated via MC simulations far from thermalization.

Thus, can we use alternative matrices differently from the ones considered
in \cite{Vinayak2014,Biswas2017}, i.e., considering the critical behavior
far from equilibrium? In addition, can we use the spectral properties to
determine the critical parameter of the spin model studied?

\begin{figure*}[t!]
\begin{center}
\includegraphics[width=2.0\columnwidth]{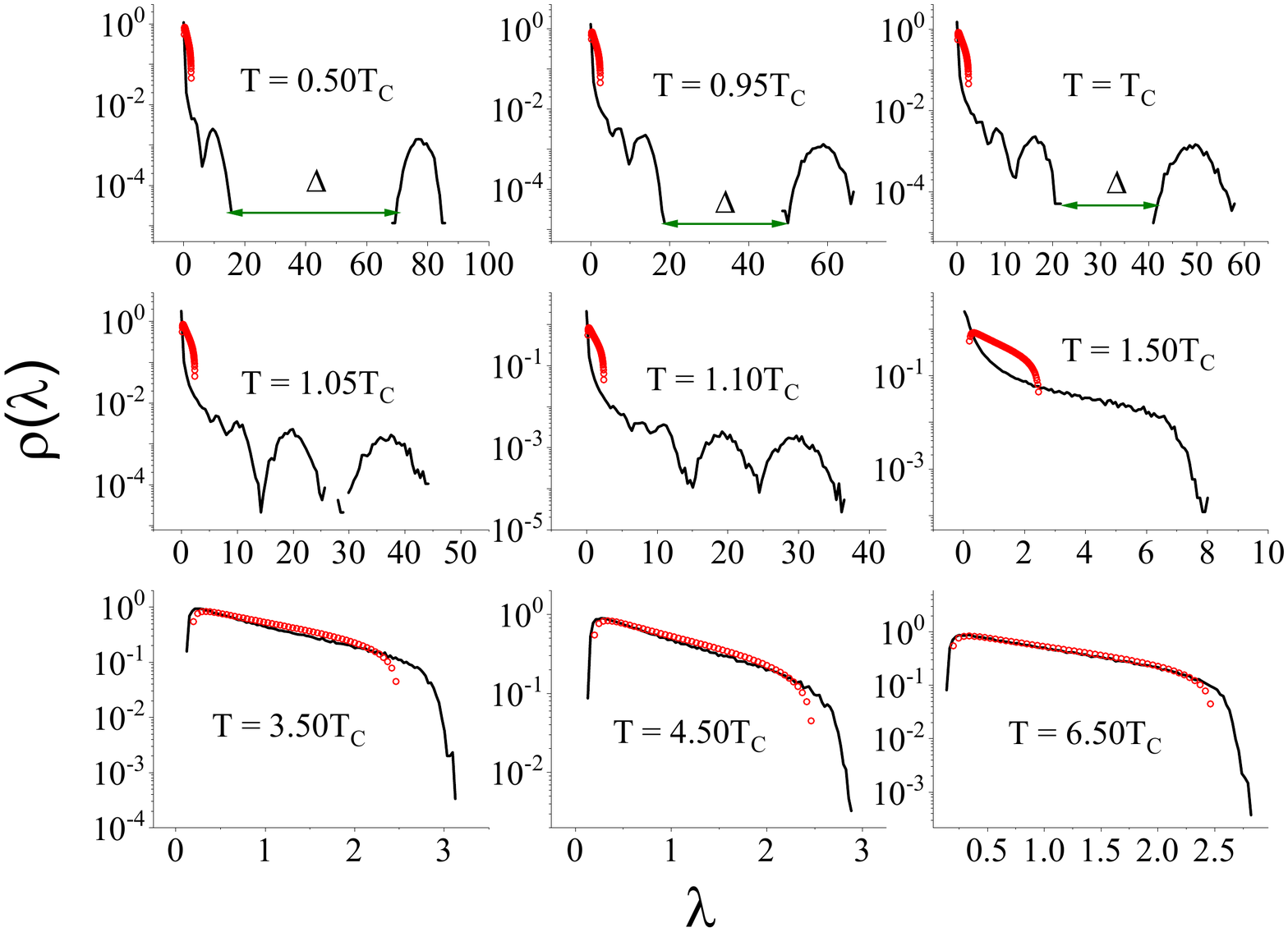}
\end{center}
\caption{The density of eigenvalues for different temperatures. Gaps of
eigenvalues disappear for $T>T_{C}$. One observes Marchenko-Pastur law when $%
T$ is large enough. }
\label{Fig:Density_diferent_frames}
\end{figure*}

Our goal in this paper is to show that it is possible. The success of our
approach is to use the correct matrix that considers magnetization time
series and not a matrix of the individual spins.

Thus, using this matrix that captures the collective character of the
system, we show that the density of eigenvalues presents an eigenvalues gap
intimately linked to the proximity of the critical system.

One performs that by first building a matrix $M$ that stores a number $%
N_{sample}$ of time series with $N_{MC}$ Monte Carlo (MC) steps. With this
in hand, we show that the density of eigenvalues of the correlation
magnetization matrix of the Ising model, built from M, presents a minimum
strictly at its critical temperature, which corroborates the inflection
point of the dispersion of eigenvalues.

In the following, we show how to define the magnetization matrix $M$ for a
correct analysis of the spectra for the localization of the critical
parameter of the Ising model in the earlier times of evolution. In the
sequence, we present our results, followed by some summaries and our
conclusions.

\begin{figure}[t!]
\begin{center}
\includegraphics[width=1.0\columnwidth]{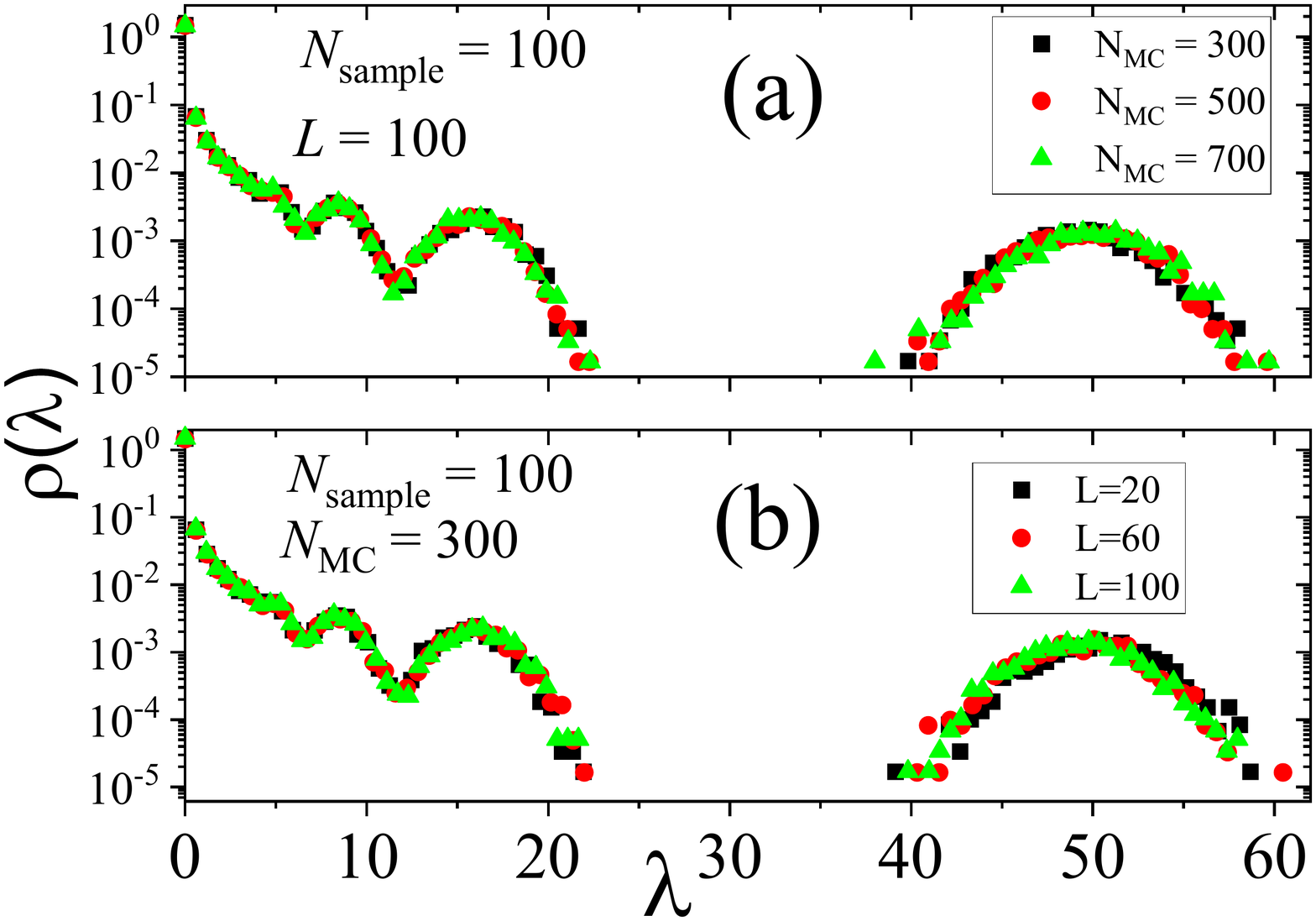}
\end{center}
\caption{Effects of the size lattice ($L$) and number of MC steps ($N_{MC}$%
). }
\label{Fig:finite_size_scaling}
\end{figure}

\textbf{2-Marcenko-Pastur's theorem and magnetization matrix}: Here we
define the main object for our analysis, the magnetization matrix element $%
m_{ij}$ that denotes the magnetization of the $j$th time series at the $i$th
MC step of a system with $N=L^{d}$ spins. For simplicity, we used $d=2$ (the
minimal dimension to appear phase transition in the simple Ising model) in
this work. Here $i=1,...,N_{MC}$, and $j=1,...,N_{sample}$. So the
magnetization matrix $M$ is $N_{MC}\times N_{sample}$. In order to analyze
spectral properties, an interesting alternative is to consider not $M$ but
the square matrix $N_{sample}\times $ $N_{sample}$:

\begin{equation*}
G=\frac{1}{N_{MC}}M^{T}M
\end{equation*}%
such that $G_{ij}=\frac{1}{N_{MC}}\sum_{k=1}^{N_{MC}}m_{ki}m_{kj}$. At this
point, instead of working with $m_{ij}$, it is more convenient to take the
Matrix $M^{\ast }$, defining its elements by the standard variables:%
\begin{equation*}
m_{ij}^{\ast }=\frac{m_{ij}-\left\langle m_{j}\right\rangle }{\sqrt{%
\left\langle m_{j}^{2}\right\rangle -\left\langle m_{j}\right\rangle ^{2}}}
\end{equation*}%
where: 
\begin{equation*}
\left\langle m_{j}^{k}\right\rangle =\frac{1}{N_{MC}}%
\sum_{i=1}^{N_{MC}}m_{ij}^{k}
\end{equation*}

Thereby: 
\begin{equation*}
\begin{array}{lll}
G_{ij}^{\ast } & = & \frac{1}{N_{MC}}\sum_{k=1}^{N_{MC}}\frac{%
m_{ki}-\left\langle m_{i}\right\rangle }{\sqrt{\left\langle
m_{i}^{2}\right\rangle -\left\langle m_{i}\right\rangle ^{2}}}\frac{%
m_{kj}-\left\langle m_{j}\right\rangle }{\sqrt{\left\langle
m_{j}^{2}\right\rangle -\left\langle m_{j}\right\rangle ^{2}}} \\ 
&  &  \\ 
& = & \frac{\left\langle m_{i}m_{j}\right\rangle -\left\langle
m_{i}\right\rangle \left\langle m_{j}\right\rangle }{\sigma _{i}\sigma _{j}}%
\end{array}%
\end{equation*}%
where $\left\langle m_{i}m_{j}\right\rangle =\frac{1}{N_{MC}}%
\sum_{k=1}^{N_{MC}}m_{ki}m_{kj}$ and $\sigma _{i}=\sqrt{\left\langle
m_{i}^{2}\right\rangle -\left\langle m_{i}\right\rangle ^{2}}$.
Analytically, if $m_{ij}^{\ast }$ are uncorrelated random variables, the
density of eigenvalues $\rho (\lambda )$ of the matrix $G=\frac{1}{N_{MC}}%
M^{\ast T}M^{\ast }$ follows the known Marcenko-Pastur distribution \cite%
{Marcenko}, which for our case we write as:

\begin{equation}
\rho (\lambda )=\left\{ 
\begin{array}{l}
\dfrac{N_{MC}}{2\pi N_{sample}}\dfrac{\sqrt{(\lambda -\lambda _{-})(\lambda
_{+}-\lambda )}}{\lambda }\ \text{if\ }\lambda _{-}\leq \lambda \leq \lambda
_{+} \\ 
\\ 
0\ \text{otherwise}%
\end{array}%
\right.   \label{Eq:MP}
\end{equation}%
where 
\begin{equation*}
\lambda _{\pm }=1+\frac{N_{sample}}{N_{MC}}\pm 2\sqrt{\frac{N_{sample}}{%
N_{MC}}}.
\end{equation*}

Sure, in the case of $m_{ij}$ is the averaged magnetization, we expect that
for $T>>T_{c}$ the density of eigenvalues $\rho ^{\exp }(\lambda )$ obtained
from computational simulations must follow $\rho (\lambda )$ in Eq. \ref%
{Eq:MP}. The question is what happens when $T\approx T_{C}$. Moreover, it
would be more interesting if the density $\rho ^{\exp }(\lambda )$, the one
obtained from computer simulations, should be used to obtain the critical
parameter of spin models.

\begin{figure*}[t!]
\begin{center}
\includegraphics[width=1.0\columnwidth]{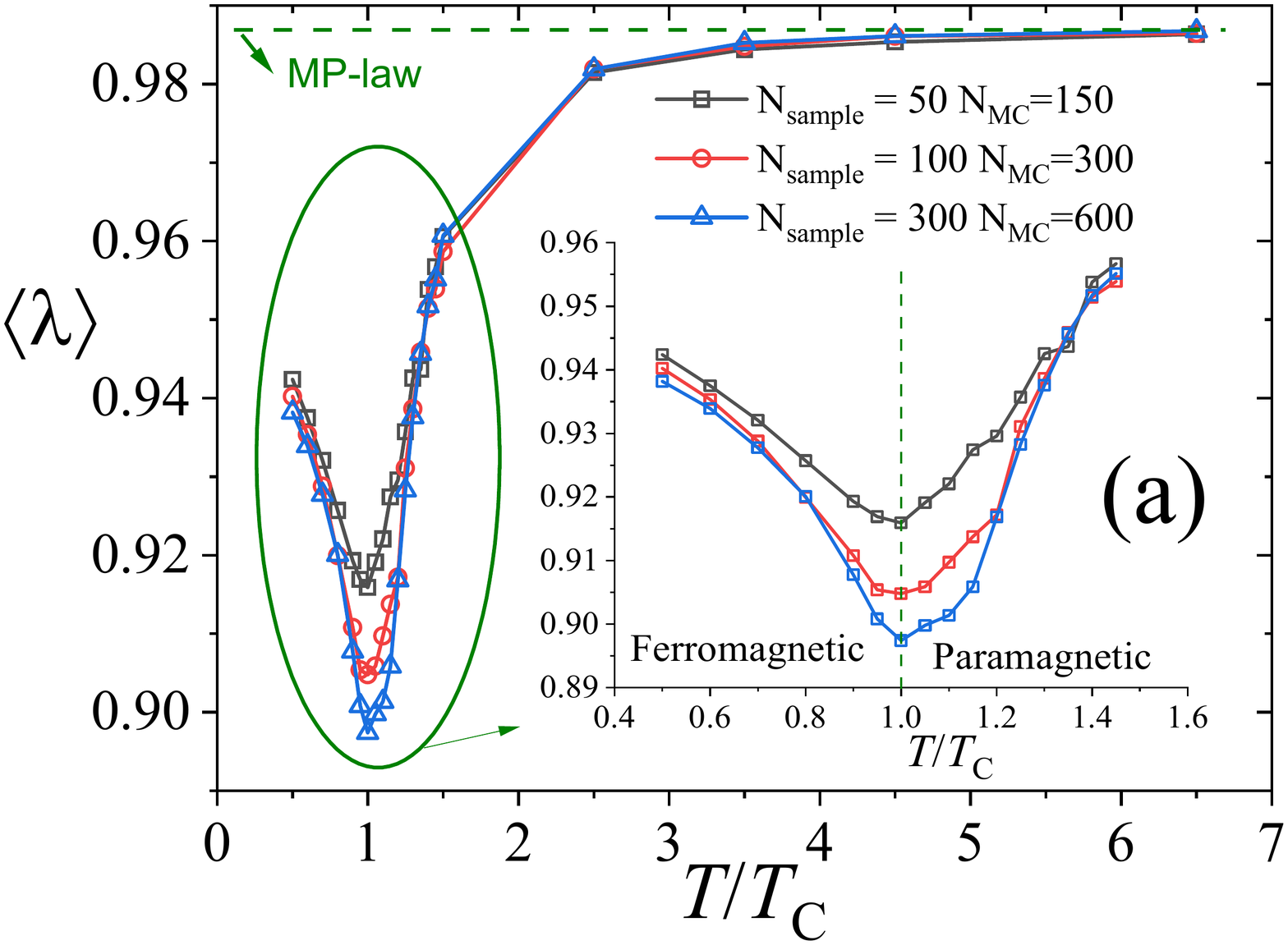} %
\includegraphics[width=1.0\columnwidth]{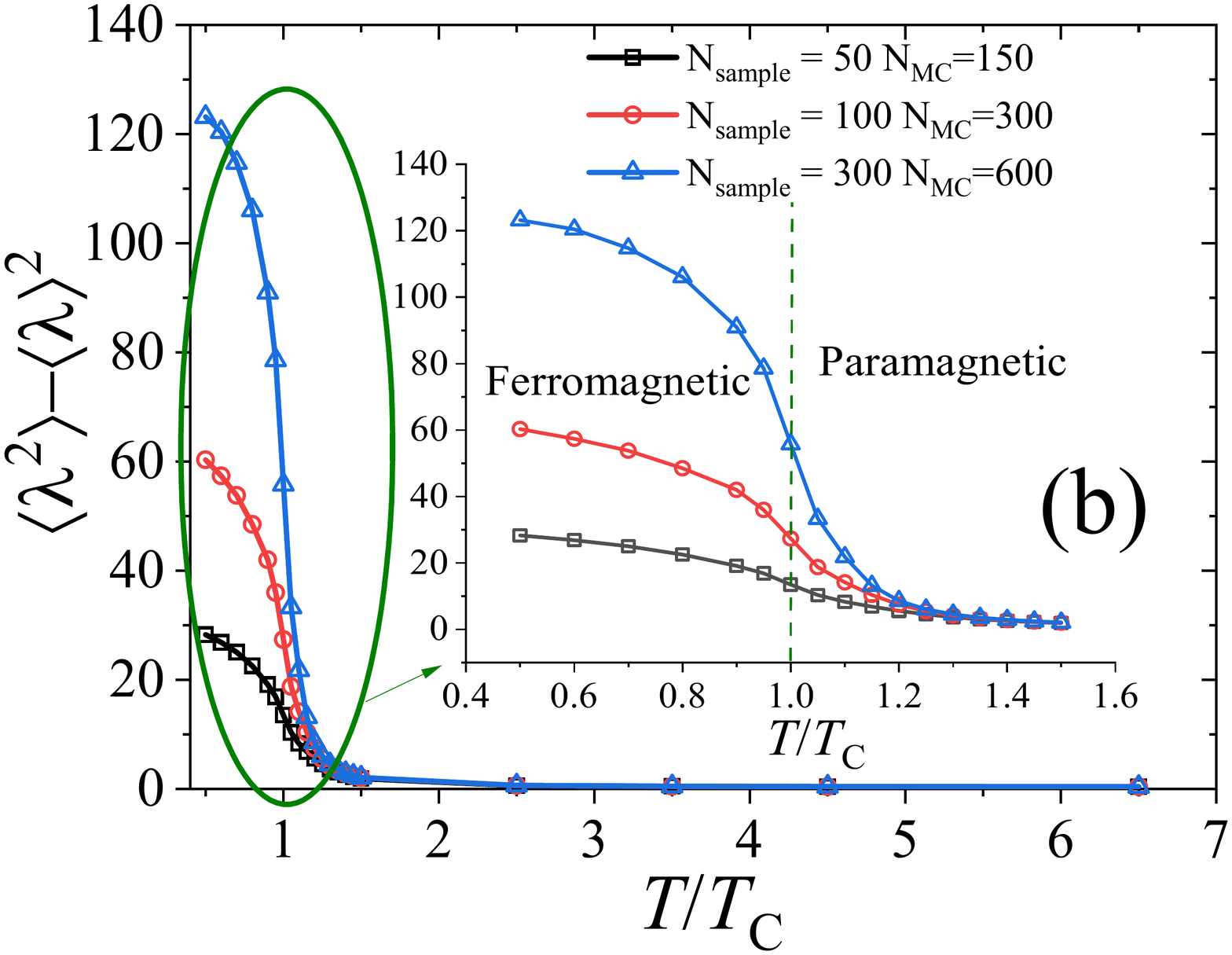}
\end{center}
\caption{(a) The minimal value of the average eigenvalue corresponds phase
transition (b) The inflection point occurs strictly at the critical
temperature. }
\label{Fig:average_variance}
\end{figure*}

\textbf{3 - Results}: We simulated different time evolutions of
magnetization of the Ising model. We used $L=100$, or $N=10^{4}$ spins in
this work, except when explicitly mentioned. Fig. \ref%
{Fig:Differenttimeevolutions} shows 20 different time series simulated at $%
T=T_{C}$ but starting from a random initial condition such that: $p(\uparrow
)=p(\downarrow )=1/2$ ($T\rightarrow \infty $).

Thus, for each temperature $T$, we simulated the Ising model building an
ensemble $N_{run}=1000$ square matrices $G^{\ast }$, built from rectangular
matrices $N_{sample}\times N_{MC}$, with $N_{sample}=100$ and $N_{MC}=300$,
except when explicitly mentioned. Thus we calculated the density of
eigenvalues $\rho ^{\exp }(\lambda )$ for each temperature as shown in Fig. %
\ref{Fig:Density_diferent_frames}.

We can observe that for $T<T_{C}$, a gap of eigenvalues characterizes the
density of eigenvalues. This gap occurs until the proximity of $T_{C}$. For $%
T=1.05T_{C}$ the gap almost disappears, which entirely happens for $%
T=1.10T_{C}$. Thus we observe a migration in the density of eigenvalues as $%
T $ increases. The Marchenko-Pastur law \ref{Eq:MP} (red points) fits the
density of eigenvalues for large T as can be observed, for example, for $%
T=6.5$ $T_{C}$.

An essential computational detail is that the density of eigenvalues seems
to be similar for the different number of MC steps as observed in Fig. \ref%
{Fig:finite_size_scaling} (a) for small systems (b).

Although the density of eigenvalues changes with temperature and the gap
after $T_{C}$ disappears, we would like to obtain a more precise parameter
to localize the critical temperature of the system quantitatively. A natural
choice is to compute the moments of the density of eigenvalues:

\begin{equation*}
\left\langle \lambda ^{k}\right\rangle =\int_{-\infty }^{\infty }\lambda
^{k}\rho _{\exp }(\lambda )d\lambda 
\end{equation*}%
where observe $\left\langle \lambda \right\rangle $ and $var(\lambda
)=\left\langle \lambda ^{2}\right\rangle -\left\langle \lambda \right\rangle
^{2}$ as function of $T$ which is shown in \ref{Fig:average_variance} (a)
and (b). We observe a notorious minimal value of $\left\langle \lambda
\right\rangle $ (Fig. \ref{Fig:average_variance} (a) ) exactly at $T=T_{C}$,
showing this amount captures the evolution of the density of eigenvalues and
the gap of the eigenvalues that appears for $T\leq T_{C}$. This minimal
seems to ocurr at $T=T_{C}$ independently on $N_{sample}$, keeping constant
the ratio $Q=\frac{N_{sample}}{N_{MC}}$. Fig. \ref{Fig:average_variance} (b)
shows that at $T=T_{C}$ we have a corresponding inflection point. \ 

\textbf{4 - Conclusions}: These results corroborate that the spectrum of
correlation matrices built from the time series of magnetization of the
Ising model in the earlier times of the evolution can precisely identify the
critical temperature of the model. The moments of the density of eigenvalues
seem to be suitable amounts to perform that. The model is promising, and it
deserves an exploration of other spin systems.

\textbf{Acknowledgements} R.d.S. thanks CNPq for financial support under
Grant No. 311236/2018-9.

\end{document}